\documentclass[11pt,openbib]{iopart}
\usepackage{iopams}
\usepackage[pdftex]{graphicx}
\pdfoutput=1

\def\be{\begin{equation}}
\def\ee{\end{equation}}
\def\bi{\begin{itemize}} 
\def\ei{\end{itemize}}
\def\ben{\begin{enumerate}}
\def\een{\end{enumerate}}

\begin{document}
\DeclareGraphicsExtensions{.pdf,.png,.gif,.jpg}

\begin{center}
\vspace{5mm}
{\Large{\bf{Coherent method for detection of gravitational wave bursts}}} \\
\vspace{5mm}
S.~Klimenko, I.~Yakushin\dag, A.~Mercer, G.~Mitselmakher \\
{University of Florida, P.O.Box 118440, Gainesville, Florida, 32611, USA} \\
{\dag LIGO Livingston observatory, P.O. Box 940, Livingston, Louisiana, 70754, USA}

\hspace{6mm}LIGO-P070093-00-Z
\end{center}

\begin{abstract}
We describe a coherent network algorithm for detection and 
reconstruction of gravitational wave bursts. 
The algorithm works for two and more arbitrarily aligned detectors and can be used for both
all-sky and triggered burst searches. We describe the main 
components of the algorithm, including the time-frequency analysis in wavelet domain, 
construction of the likelihood time-frequency maps, the identification and selection of 
burst events. 
\end{abstract}


\section{Introduction}

Coherent network analysis is addressing the problem of detection and reconstruction
of gravitational waves (GW) with networks of detectors. 
It has been extensively studied in the literature \cite{GT,FH,Arnaud,prd,Malik,cit-na}
in application to detection of bursts signals, which may be produced 
by numerous gravitational wave sources in the
Universe~\cite{SN1,SN2,SN3,SN4,BH_FH,BH_NASA,BH_P,BH_UTB,GRB}.
In coherent methods, a statistic is built up as a coherent sum over detector responses.
In general, it is expected to be more optimal (better sensitivity at the same false alarm rate) 
than the detection statistics of the individual detectors that make up the network. 
Also coherent methods provide estimators for the GW waveforms and the source coordinates 
in the sky.

The method we present (called coherent WaveBurst) is significantly different from the traditional 
burst detection methods.
Unlike coincident methods \cite{WB1,Q,BN}, which first identify 
events in individual detectors by using an excess power statistic and than require coincidence 
between detectors, the coherent WaveBurst method combines all data streams into one coherent statistic 
constructed in the framework of the constrained maximum likelihood analysis \cite{prd}. 
Such an approach 
has significant advantages over the coincident methods. First, the sensitivity 
of the method is not limited by the least sensitive detector in the network. 
In the coherent WaveBurst method the 
detection is based on the maximum likelihood ratio statistic which represents 
the total signal-to-noise 
ratio of the GW signal detected in the network. Second, other coherent statistics, 
such as the null stream and the network correlation coefficient can be constructed 
to distinguish genuine GW signals from the environmental and instrumental artifacts. 
Finally, the source coordinates of the GW waveforms can be reconstructed. 

\section{Coherent analysis}
\label{CL}

The coherent WaveBurst pipeline (cWB) uses a method, for a coherent detection and 
reconstruction of burst signals, based on the use of the likelihood ratio functional~\cite{prd}.
For a general case of Gaussian quasi-stationary noise it can be written in the wavelet 
(time-frequency) domain as
\begin{eqnarray}
\label{eq:like}
{\cal{L}} = \sum_{k=1}^K \sum_{i,j=1}^{N} 
 \left( \frac{w^2_k[i,j]}{\sigma^2_k[i,j]}  - 
\frac{\left(w_k[i,j] - \xi_k[i,j]\right)^2}{\sigma^2_k[i,j]} \right) \;,  
\end{eqnarray}
where $K$ is the number of detectors in the network, $w_k[i,j]$ is the sampled detector data 
(time $i$ and frequency $j$ indices  run over some time-frequency area of size $N$) 
and $\xi_k{[i,j]}$ are the detector responses. Note, we omit the term $1/2$ in
the conventional definition of the likelihood ratio.    
The detector noise is characterised by its standard deviation $\sigma_k[i,j]$, which
may vary over the time-frequency plane. 
The detector responses are written in the standard notations
\begin{eqnarray}
\label{eq:resp}
\xi_k{[i,j]} = F_{+k}h_+[i,j] +  F_{\times{k}}h_{\times}[i,j] \;,
\end{eqnarray}
where $F_{+k}(\theta,\phi)$, $F_{\times{k}}(\theta,\phi)$ are the
detector antenna patterns  (depend
upon source coordinates $\theta$ and $\phi$) and $h_+[i,j]$,  
$h_{\times}[i,j]$ are the two polarisations of the gravitational wave signal
in the wave frame. Since the detector responses $\xi_k$ are invariant with respect to the
rotation around z-axis in the wave frame, the polarization angle is included in the definition
of $h_+$ and $h_{\times}$. 
The GW waveforms, $h_+$ and $h_{\times}$, are found by variation of ${\cal{L}}$.
The maximum likelihood ratio statistic is obtained by substitution of the solutions 
into the functional ${\cal{L}}$.  The waveforms in the time domain are reconstructed from 
the inverse wavelet transformation.
Below, for convenience we introduce the data vector ${\bf{w}}[i,j]$ and the antenna pattern vectors
${\bf{f_+}}[i,j]$ and ${\bf{f_\times}}[i,j]$
\begin{eqnarray}
\label{eq:vec1}
{\bf{w}}[i,j] = \left( \frac{w_1[i,j]}{\sigma_1[i,j]},..,\frac{w_K[i,j]}{\sigma_K[i,j]} \right) \; \\
\label{eq:vec2}
{\bf{f_{+(\times)}}}[i,j] = \left( \frac{F_{1+(\times)}}{\sigma_1[i,j]},..,\frac{F_{K+(\times)}}{\sigma_K[i,j]} \right) \;
\end{eqnarray}
Further in the text we omit the time-frequency indices and replace 
the sum $\sum_{i,j=1}^{N}$ with $\sum_{\Omega_{TF}}$, where $\Omega_{TF}$ is the time-frequency area
selected for the analysis.

The likelihood functional (Eq.\ref{eq:like}) can be written in the form
${\cal{L}}={\cal{L}}_++{\cal{L}}_\times$:
\begin{eqnarray}
\label{eq:like1}
{\cal{L}}_+ =  \sum_{\Omega_{TF}} \left[ ({\bf{w}}\cdot{\bf{f_+}}) h_+ - \frac{1}{2}|{\bf{f_+}}|^2 h^2_+ \right]  \;, \\
\label{eq:like2}
{\cal{L}}_\times =  \sum_{\Omega_{TF}} \left[ ({\bf{w}}\cdot{\bf{f_\times}}) h_\times - \frac{1}{2}|{\bf{f_\times}}|^2 h^2_\times \right]  \;, 
\end{eqnarray}
where  the antenna pattern vectors ${\bf{f_+}}$ and ${\bf{f_{\times}}}$ are defined 
in the Dominant Polarisation wave Frame (DPF)~\cite{prd}. 
In this frame the antenna pattern vectors are orthogonal to each other: $(\bf{f_+}\cdot\bf{f_\times})=0$.
The estimators of the GW waveforms are the solutions of the equations
\begin{eqnarray}
\label{eq:syst1}
 {({\bf{w}}\cdot{\bf{f_+}})}  = |{\bf{f_+}}|^2 h_+ \;, \\
\label{eq:syst2}
 {({\bf{w}}\cdot{\bf{f_\times}})}  = |{\bf{f_\times}}|^2 h_\times \;. 
\end{eqnarray}
Note, the norms $|{\bf{f_+}}|^2$ and $|{\bf{f_{\times}}}|^2$ characterise the network 
sensitivity to the $h_+$ and $h_{\times}$ polarisations respectively.

\subsection{Likelihood regulators}

As first shown in~\cite{prd}, there is a specific class of constraints (often called regulators), 
which arise from the way the network responds to a generic GW signal. 
A classical example is a network of aligned detectors where the detector responses 
$\xi_k$  are identical. In this case the algorithm can be constrained to search for an unknown 
function $\xi$ rather than for two GW polarisations
$h_+$ and $h_\times$, which span a much larger parameter space. Note, that in this case 
$|{\bf{f_\times}}|^2=0$, the Equation~\ref{eq:syst2} is ill-conditioned and the solution for the 
$h_\times$ waveform can not be found.  
The regulators are important not only for aligned detectors, but also for networks 
of miss-aligned detectors, for example, the LIGO and Virgo 
network~\cite{LIGO,Virgo}.
Depending on the source location, the network can be much less sensitive 
to the second GW component ($|{\bf{f_\times}}|^2<<|{\bf{f_+}}|^2$) and the 
$h_\times$ waveform may not be reconstructed from the noisy data.

In the coherent WaveBurst analysis we introduce  a regulator by changing the norm of the 
${\bf{f_\times}}$ vector
\begin{eqnarray}
\label{eq:reg}
|{\bf{f'_\times}}|^2 = |{\bf{f_\times}}|^2 + \delta, 
\end{eqnarray}
where $\delta$ is a parameter. This is equivalent to adding one more dummy detector to the
network with the antenna patterns ${f_{+,K+1}=0}$, ${f_{\times,K+1}=\sqrt{\delta}}$ and zero detector 
output ($x_{K+1}=0$). In this case, the regulator preserves the orthogonality of the
vectors ${\bf{f_+}}$ and ${\bf{f'_\times}}$ and the maximum likelihood statistic is written as   
\begin{equation}
\label{eq:lMax}
L_{\mathrm{max}} = \sum_{\Omega_{TF}} \left [ \frac{({\bf{w}}\cdot{\bf{f_+}})^2}{|{\bf{f_+}}|^2}+
\frac{({\bf{w}}\cdot{\bf{f'_{\times}}})^2}{|{\bf{f'_{\times}}}|^2} \right ] = 
\sum_{\Omega_{TF}} \left [ ({\bf{w}}\cdot{\bf{e_+}})^2+({\bf{w}}\cdot{\bf{e'_{\times}}})^2 \right ],
\end{equation}
where the ${\bf{e_+}}$ and ${\bf{e'_{\times}}}$ are unit vectors.
Depending on the value of the parameter $\delta$ different statistics can be generated, for example:
\begin{itemize} 
\item $\delta=0$ - standard likelihood,
\item $\delta=\infty$ - hard constraint likelihood.
\end{itemize}

\subsection{Reconstruction of GW waveforms}
\label{WF}

The GW waveforms are given by the solutions of the likelihood functional
~Eq.\ref{eq:like1},\ref{eq:like2}. 
For the first GW component the solution is
\begin{eqnarray}
\label{lmax}
h_+ = \frac{({\bf{w}}\cdot{\bf{f_+}})}{|{\bf{f_+}}|^2}.
\end{eqnarray}
When the regulator is introduced it affects the solution for the second GW component. 
In this case we look for such a solution, which
gives the second term of the likelihood statistic $L_{\mathrm{max}}$, when the solution is substituted 
into the likelihood functional. Namely, we solve the equation
\begin{eqnarray}
\label{lmax2}
2 ({\bf{w}}\cdot{\bf{f_{\times}}}) h_\times - {|{\bf{f_{\times}}}|^2} h^2_\times - \frac{({\bf{w}}\cdot{\bf{f'_{\times}}})^2}{|{\bf{f'_{\times}}}|^2}=0.
\end{eqnarray}
Out of two possible solutions the following one is selected
\begin{eqnarray}
\label{lmax1}
h_\times = \frac{({\bf{w}}\cdot{\bf{f_{\times}}})}{|{\bf{f'_{\times}}}|^2} \left(1 + \sqrt{1-\frac{|{\bf{f_{\times}}}|^2}{|{\bf{f'_{\times}}}|^2}} \right)^{-1}.
\end{eqnarray}
In case of aligned detectors ($|f_\times|=0$) this equation results in a trivial solution $h_\times=0$.

\section{Data analysis algorithms}
\label{algorithms}

In this section we describe the algorithms used in the coherent WaveBurst 
pipeline. They include: wavelet transformation, conditioning of input data, construction
of time delay filters, and generation of coherent triggers. 

\subsection{Wavelet transformation}
\label{tree}

The discrete wavelet transformations (DWT) are applied to discrete 
data and produce discrete wavelet series $w[i,j]$, where $j$ is the scale 
index and $i$ is a time index.
Applied to time series, the DWT maps data from time domain to the wavelet domain. 
All DWTs used in cWB have {\it critical sampling} when the output data 
vector has the same size as the input data vector. 

Wavelet series give a time-scale representation of data where each wavelet scale 
can be associated with a certain frequency band of the initial time series. Therefore 
the wavelet time-scale spectra can be displayed as a time-frequency (TF)
scallogram, where the scale is replaced with the central frequency $f$ of the band. 
The time series sampling rate $R$ and the scale number $j$ determine 
the time resolution $\Delta{t_j}(R)$ at this scale. The DWT preserves the 
time-frequency volume of the data samples, which is equal to $1/2$ for the input 
time series.  Therefore the frequency resolution $\Delta{f_j}$ is defined 
as $1/(2\Delta{t_j})$ and determines the data bandwidth at the scale $j$.
For optimal localisation of the GW energy on the TF plane, the cWB analysis is performed 
at several time-frequency resolutions. 

\begin{figure}[htbp]
  \begin{center}
  \resizebox{\textwidth}{10.0cm}{\rotatebox{0}{\includegraphics{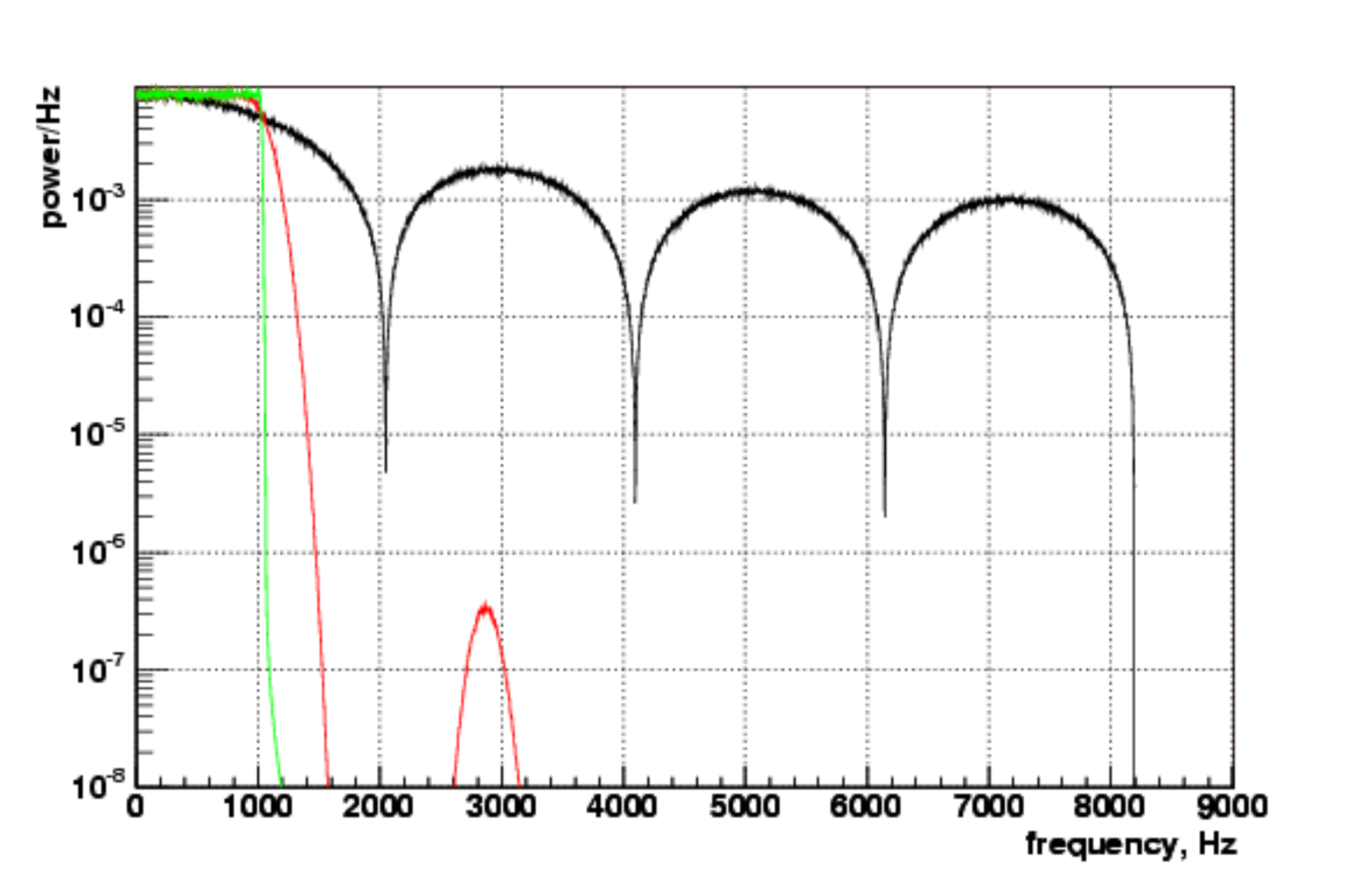}}}
  \end{center}
  \caption{Comparison of spectral leakage from the first (low) frequency band $0-1024$Hz to the high frequency
bands for Haar (black), Symlet60 (red) and Meyer1024 (green) wavelets after three wavelet decomposition steps.}
  \label{fig:haarSymMey_res3log}
\end{figure}

The time-frequency resolution defined above should be distinguished from
the intrinsic time-frequency resolution of the wavelet transformation, 
which defines the spectral leakage between the wavelet sub-bands and depends on 
the length of the wavelet filter.                          
To reduce spectral leakage we use Meyers wavelets for which long filters can be
easily constructed \cite{vidakovic}. As shown in Figure~\ref{fig:haarSymMey_res3log},
it allows us much better localization of 
the burst energy on the time-frequency plane than Symlet60 wavelets used for 
the LIGO S2-S4 analysis \cite{S2,S4}. 
The disadvantage of the Meyer filters is that for the local support they have to be truncated. 
As the result, the Meyer wavelets are approximately orthonormal. 
From the other side the Meyer filters can be constructed so that                          
the Parseval identity holds with better then $0.01\%$ accuracy, which is more than        
adequate for the analysis.                         


\subsection{Linear prediction error filter}

The linear prediction error (LPE) filters are used to remove ``predictable'' components                
from an input time series. Usually they are constructed and applied in the time domain. 
In this case the output of the LPE filter is a whitened time series. 
The LPE filters can be also used in the wavelet domain. 
For construction of the LPE filters we follow the approach described in \cite{genin87}.   
The symmetric LPE filters can be constructed from the backward and forward LPE       
filters by using classical Levinson algorithm, or the split lattice algorithm.               
               
\begin{figure}[htbp]
\resizebox{\textwidth}{8.0cm}{\rotatebox{0}{\includegraphics{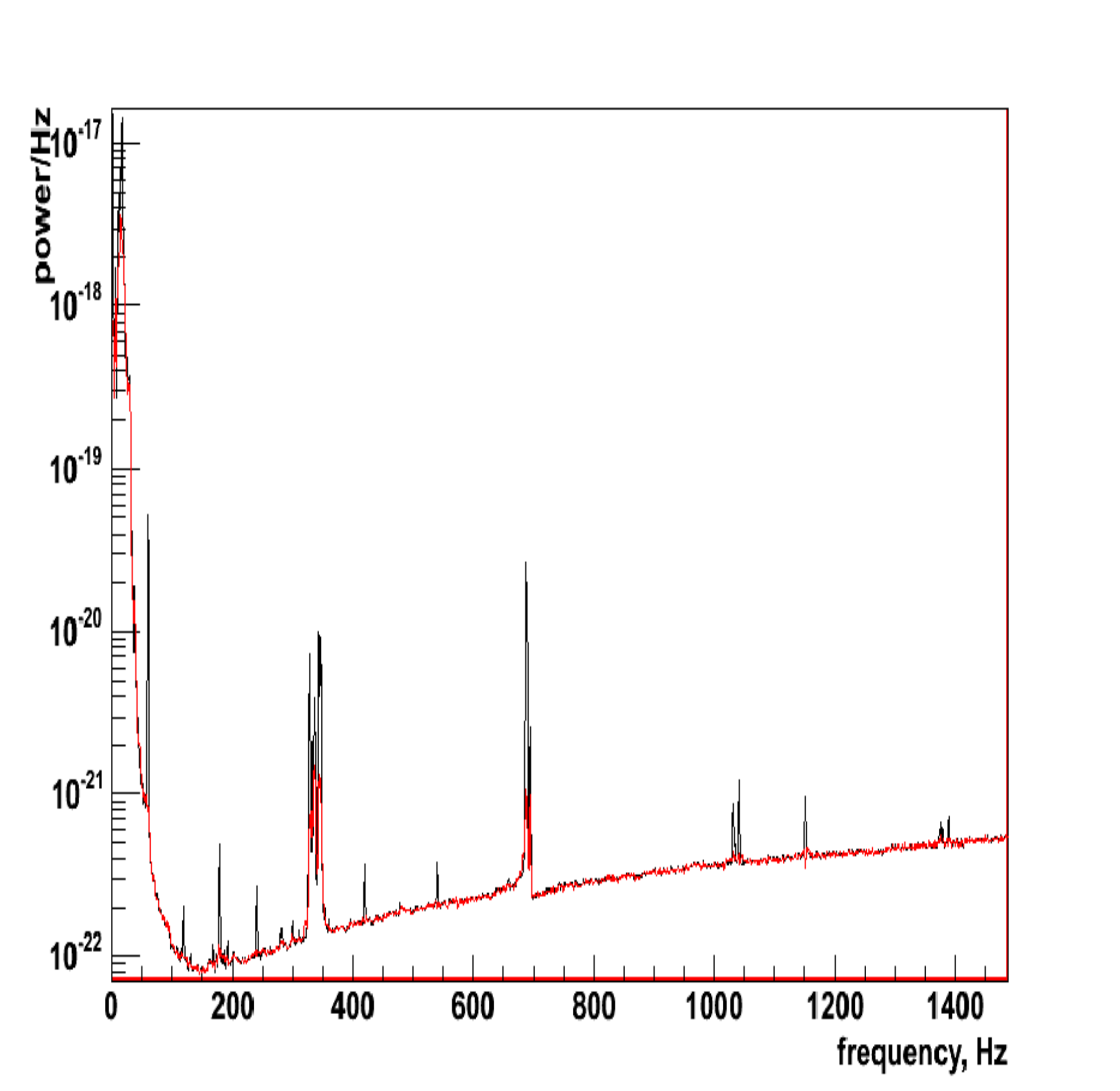}}}
  \caption{Power spectra of original (black) and LPE filtered (red) noise of the Hanford 4k detector.}
  \label{fig:watLPR_S4psd}
\end{figure}

Since each wavelet layer is a time series, rather then applying the LPE filter to a time               
series $x(t)$, one can perform a wavelet decomposition $x(t) \rightarrow w(f,t)$ first, and  
then construct and apply the LPE filter $F(f)$ individually to each wavelet layer.                
A set of filters $F(f)$ removes predictable components (like lines) in the wavelet layers               
producing data $w'(t)$. The filtered time series $x'(t)$ can be reconstructed from $w'(t)$ with the inverse wavelet transformation. 
An example PSD of the filtered segment of S4 data is shown in Figure~\ref{fig:watLPR_S4psd}. 
As one can see, when applied in wavelet domain the LPE filter removes spectral lines               
but preserves the power spectral density of the noise floor.

\subsection{Time delay filters in the wavelet domain}
\label{delay}

The likelihood method requires calculation of the inner products 
$\left<x_n(\tau_n),x_m(\tau_m)\right>$, where the
data streams are shifted in time to take into account the GW signal time delay
between the detectors $n$ and $m$. The time delay $\tau_{n}-\tau_{m}$ in turn, 
depends on the source coordinates $\theta$ and $\phi$.

In time domain it is straightforward to account for the time delays.
However for colored detector noise it is preferable to 
calculate the maximum likelihood statistics in the Fourier or 
wavelet (time-frequency) domains. 
In the wavelet domain one needs to calculate the inner products
$\left<w_n(\tau_n),w_m(\tau_m)\right>$. 
The delayed amplitudes can be calculated from the original amplitudes 
(before delay) with the help of the time delay filter $D_{kl}(\tau)$
\begin{equation}\label{eq:rsig}
   w_{n,m}(i,j,\tau) = \sum_{kl}{D_{kl}(\tau,j)w_{n,m}(i+k,j+l)},
\end{equation}
where $k$ and $l$ are the local TF coordinates with respect to the TF 
location ($i,j$). The delay filters are constructed individually for 
each wavelet layer, which is indicated with the index $j$.

The construction of time delay filters is related to the decomposition 
of the sampled wavelet functions $\Psi_j(t+\tau)$ in the basis of the non-shifted
wavelet functions $\Psi_j(t)$. The filter construction procedure
can be described in the following steps:
\begin{itemize}
\item create a wavelet series with only one coefficient at the
      TF location (i,j) set to unity,
\item apply the inverse wavelet transformation reconstructing
      $\Psi_j(t)$ in time domain,
\item shift $\Psi_j(t)$ by delay time $\tau$ and perform wavelet decomposition of $\Psi_j(t+\tau)$,
\item the resulting wavelet amplitudes at the TF locations $(i+k,j+l)$ give the delay 
      filter coefficients $D_{kl}(\tau,j)$ for the wavelet layer $j$.
\end{itemize}

The length of the filter is determined by the requirement on the
acceptable energy loss when the time delay filter is applied. 
The fractional energy loss is
\begin{equation}\label{eq:rsig1}
   \epsilon_K = 1-\sum_{K}{D^2_{kl}},
\end{equation}
where the sum is calculated over the $K$ most significant coefficients.
The selected coefficients are also described by the list of their relative TF
locations $(k,l)$ which should be stored along with the filter coefficients
$D_{kl}$.
Typically $K$ should be greater then 20 to obtain the fractional energy loss 
less than 1\%. 

\subsection{Generation of coherent triggers}
\label{ctrig}

A starting point of any burst analysis is the identification of burst events (triggers). 
Respectively, this stage of the burst analysis pipeline is called
the event trigger generator (ETG). Usually, the ETGs based on the excess power 
statistics of individual detectors are used in the analyses \cite{WB1,Q,BN}.
Another example of an ETG is the CorrPower algorithm \cite{r-stat}, which uses cross-correlation
between aligned detector pairs to generate the triggers. 
The likelihood statistic used in the coherent WaveBurst utilizes both the excess power and 
the cross-correlation terms. 

\subsubsection{Likelihood time-frequency maps}
\label{TFmap}

\begin{figure}[htb]
\resizebox{\textwidth}{10.0cm}{\rotatebox{0}{\includegraphics{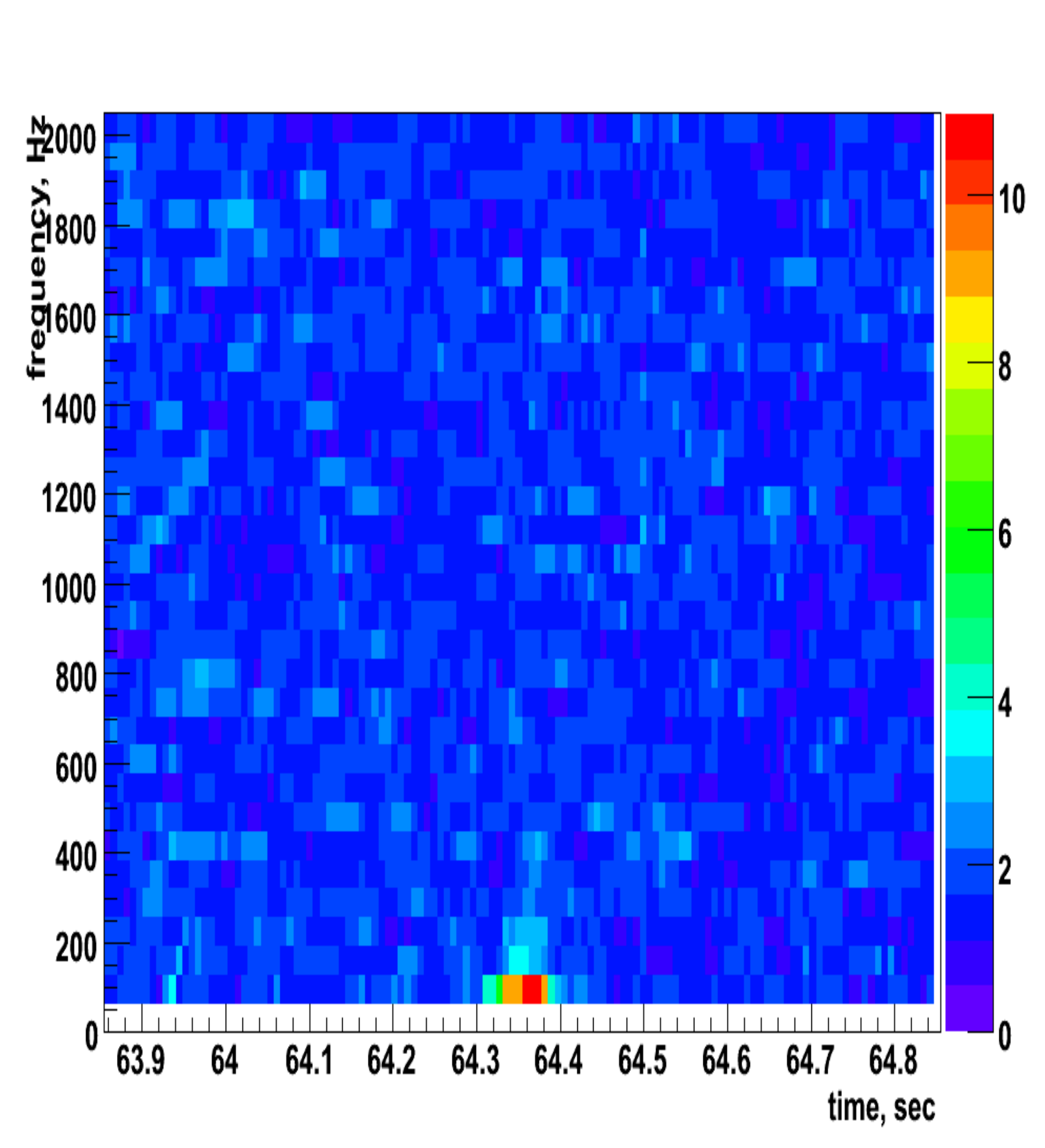}}}
   \caption{Example of the likelihood time-frequency map for a magnetic glitch in the S4 L1xH1xH2 data.}
   \label{fig:LTFmap}
\end{figure}

In general the likelihood functional is calculated as a sum over
the data samples selected for the analysis (see Eq.\ref{eq:like}).
The number of terms in the sum depends on the selected TF area in 
the wavelet domain. When the sum consists of 
only one term, one can write the likelihood functional for a given
TF location and point in the sky~\footnote{For definition of vectors ${\bf{w}}$, ${\bf{f_+}}$, and ${\bf{f_\times}}$ 
see Eq.\ref{eq:vec1},\ref{eq:vec2}}:
\begin{equation}\label{eq:ltf}
   {\cal{L}}_p(i,j,\theta,\phi) = {|{\bf{w}}|^2} - |{\bf{w}}-{\bf{f_+}}h_+-{\bf{f_\times}}h_\times|^2.  
\end{equation}
Since the entire likelihood approach is applicable to the functional above, one can solve
the variation problem and find the maximum likelihood statistics $L_p(\theta,\phi)$.
They can be maximized over the source coordinates $\theta$ and $\phi$, resulting in the statistics
\begin{equation}\label{eq:lmm}
   L_{m}(i,j) = {\mathrm{max}}_{\theta,\phi}\{L_p(i,j,\theta,\phi)\}.
\end{equation}
Calculated as a function of time and frequency, it gives a likelihood
time-frequency (LTF) map. Figure~\ref{fig:LTFmap} shows an example of the LTF map for 
a segment of the S4 data. 

A single data sample in the map is called the LTF pixel.  It is characterized by its TF location ($i,j$)
and by the arrays of wavelet amplitudes $w_k(i,j,\tau_{k}(\theta,\phi))$, which are used to
construct the likelihood statistics $L_{p}$. 

\subsubsection{Coherent triggers}
\label{trigger}

The statistic $L_{m}$ has a meaning of the maximum possible energy detected 
by the network at a given TF location. By selecting the values of $L_{m}$ 
above some threshold, one can identify groups of the LTF pixels 
(coherent trigger) on the time-frequency plane.
A coherent trigger is defined for the entire network, rather 
than for the individual detectors. Therefore, further in the text we reserve 
a name ``cluster'' for a group of pixels selected in a single detector 
and refer to a group of the LTF pixels as a coherent or network trigger.

After the coherent triggers are identified, one has to reconstruct
the parameters of the GW bursts associated with the triggers, including the 
reconstruction of the source coordinates, the two GW polarisations, the individual detector 
responses and the maximum likelihood statistics of the triggers. The likelihood of 
reconstructed triggers is calculated as
\begin{equation}\label{eq:LTRIG}
   {\cal{L}}_{c}(\theta,\phi) = \sum_{ij}{{\cal{L}}_p(i,j,\theta,\phi)}
\end{equation}
where the sum is taken over the LTF pixels in the trigger.
The maximum likelihood statistic $L_{\mathrm{max}}$ is obtained by variation 
of $L_c$ over $\theta$ and $\phi$. Unlike for $L_p$, which is calculated 
for a single LTF pixel, the $L_{\mathrm{max}}$ is calculated simultaneously 
for all LTF pixels forming the coherent trigger.

%
%
%


\section{Selection of coherent triggers}
\label{post}
When the detector noise is Gaussian and stationary, the maximum likelihood $L_{\mathrm{max}}$ is the
only statistic required for detection and selection of the GW events. In this case
the false alarm and the false dismissal probabilities are controlled by the threshold 
on $L_{\mathrm{max}}$. The real data however, is contaminated with the instrumental and 
environmental glitches and additional selection cuts should be applied to distinguish 
genuine GW signals~\cite{cit-na,nullAEI}. 
Such selection cuts test the consistency of the reconstructed responses in the detectors.  
In the coherent WaveBurst method the consistency test is based on the 
coherent statistics constructed from the elements of the likelihood and the null matrices.


The likelihood matrix $L_{nm}$ is obtained from the likelihood quadratic form (see Eq.\ref{eq:lMax})
%
\begin{equation}\label{eq:lMax1}
   L_{\mathrm{max}} = \sum_{nm}{L_{nm}} = \sum_{nm}{\left[\left<w_nw_me_{+n}e_{+m}\right>+
   \left<w_nw_me'_{{\times}n}e'_{{\times}m}\right>\right]}. 
\end{equation}
where $n$ and $m$ are the detector indexes.
The diagonal (off-diagonal) terms of the matrix 
$L_{mn}$ describe the reconstructed normalized incoherent (coherent) energy.
The sum of the off-diagonal terms is the coherent energy $E_{\mathrm{coh}}$ detected by the network.
The coherent terms can be used to construct the correlation 
coefficients:
\begin{equation}\label{eq:rij}
   r_{nm} = \frac{L_{nm}}{\sqrt{L_{nn}L_{mm}}}.
\end{equation}
which represent Pearson's correlation coefficients in the case of aligned detectors.
We use the coefficients $r_{nm}$ to construct the reduced coherent energy
\begin{equation}\label{eq:rij1}
   e_{\mathrm{coh}} = \sum_{nm}{L_{nm}|r_{nm}|},~~n\neq{m}.
\end{equation}
which provides one of the most efficient selection cuts for rejection of the
incoherent background events. 
 
The null matrix represents the normalized energy of the reconstructed noise
\begin{equation}\label{eq:Nmatrix}
   N_{nm} = E_{nm} - L_{nm},
\end{equation}
where $E_{nm}$ is the diagonal matrix of the normalized energy in the detectors:
$E_{nn}=\left<x^2_n\right>$.
To distinguish genuine GW signals from the instrumental and invironmental glitches
we introduce the network correlation coefficients
\begin{equation}\label{eq:ltf1}
   C_{\mathrm{net}} = \frac{E_{\mathrm{coh}}}{N_{\mathrm{ull}}+|E_{\mathrm{coh}}|},~~c_{\mathrm{net}} = \frac{e_{\mathrm{coh}}}{N_{\mathrm{ull}}+|e_{\mathrm{coh}}|} 
\end{equation}
where $N_{\mathrm{ull}}$ is a sum of all elements of the null matrix, which represents the total energy
in the null stream.
Usually for glitches little coherent energy is detected and
the reconstructed detector responses are inconsistent with the detector outputs which results
in the large null energy. Therefore the correlation coefficients $C_{\mathrm{net}}$ and $c_{\mathrm{net}}$
can be used for a signal consistency test which effectively compares the null energy with the coherent energy. 
This is much safer consistency test than the
null stream veto~\cite{nullAEI} where the null energy is compared with the estimated noise energy.
In any realistic data analysis there is always some residual energy left in the
null stream. Therefore for strong gravitational waves the energy of the residual signal 
can be much larger than the noise energy that may result in the false rejection of the GW signals.
This is not the case for the veto cut based on the $C_{\mathrm{net}}$ and $c_{\mathrm{net}}$.

\section{Summary}
In the paper we discussed how the coherent network algorithms are constructed for burst
searches. We found it convenient to construct coherent burst searches in the time-frequency
(wavelet) domain, which requires construction of time delay filters. For detection
of burst signals  we combine output of all detectors into one coherent 
statistic - likelihood, which represents the total signal-to-noise ratio of the signal
detected in the network. To distinguish genuine GW signals from the instrumental and 
environmental glitches we introduced several coherent statistics constructed from the
elements of the likelihood and null matrices. We do not discuss the performance of the method
in this paper, however, numerous studies of the method with different sets
of LIGO and Virgo data have been performed. It was found, in general, that the method has better performance
than the burst algorithm used for the published analysis of the LIGO data~\cite{WB1,LIGO,Virgo}. 
The results of these studies will be presented in subsequent papers.

\section{Acknowledgments}
We thank Keith Riles, Michele Zanolin and Brian O'Reilly for detailed discussions
and review of the algorithm and suggestions which significantly improved its performance.
This work was supported by the US National Science Foundation grant PHY-0555453 
to the University of Florida, Gainesville.


\end{document}